\documentclass[aps,prl,letterpaper,twocolumn,nofootinbib,floatfix,showpacs]{revtex4}

\usepackage{amsfonts}
\usepackage{amsmath}
\usepackage{amssymb}
\usepackage{graphicx}

\begin{document}



\title{Fast DNA sequencing via transverse electronic transport}

\renewcommand{\thefootnote}{\fnsymbol{footnote}}
\author{$^{1}$Johan Lagerqvist, $^{2}$Michael Zwolak, and $^{1}$Massimiliano Di
Ventra\cite{MD}}
\affiliation{$^{1}$Department of Physics, University of California, San Diego, La Jolla,
CA 92093-0319}
\affiliation{$^{2}$Physics Department, California Institute of Technology, Pasadena,
CA 91125}
\date{\today}
\begin{abstract}
A rapid and low-cost method to sequence DNA would usher in a revolution in medicine. We propose and theoretically show the feasibility of a protocol for sequencing based on the distributions of transverse electrical currents of single-stranded DNA while it translocates through a nanopore.  Our estimates, based on the statistics of these distributions, reveal that sequencing of an entire human genome could be done with very high accuracy in a matter of hours without parallelization, e.g., orders of magnitude faster than present techniques. The practical implementation of our approach would represent a substantial advancement in our ability to study, predict and cure diseases from the perspective of the genetic makeup of each individual.
\end{abstract}

\maketitle
Recent innovations in manufacturing processes have made it possible to fabricate devices with pores at the nanometer scale~\cite{LiNat01,StormNatmat03,HarrellAnalchem03,LiAnalchem04,LemayAnalchem05}, i.e., the scale of individual nucleotides. This opens up fascinating new venues for sequencing DNA. For instance, one suggested method is to measure the so called blockade current~\cite{KasianowiczProcusa96,AkesonBiophys99,DeamerTrends00,MellerProcusa00,VercoutereNatbio01,MellerPRL01,DeamerAccchem02,MellerElectro02,LiNatmat03,NakaneJphyscm03,AksimentievBiophys04,ChenNanolett04,FologeaNanolett051,HengBiophys06}. In this method, a longitudinal electric field is applied to pull DNA through a pore. As the DNA goes through, a significant fraction of ions is blocked from simultaneously entering the pore. By continuously measuring the ionic current, single molecules of DNA can be detected. Other methods using different detection schemes, ranging from optical~\cite{RefAMIT} to capacitive~\cite{RefSchulten06}, have also been suggested.  Despite much effort, however, single nucleotide resolution has not yet been achieved~\cite{ChanMutat05}.

In this Letter, we explore an alternative idea which would allow single-base resolution by measuring the electrical current perpendicular to the DNA backbone while a single strand immersed in a solution translocates through a pore. To do this, we envision embedding electrodes in the walls of a nanopore as schematically shown in the inset of Figure 1. The realization of such a configuration, while difficult to achieve in practice, is within reach of present experimental capabilities~\cite{LiNat01,StormNatmat03,HarrellAnalchem03,LiAnalchem04,LemayAnalchem05}. The DNA is sequenced by using the measured current as an electronic signature of the bases as they pass through the pore. We couple molecular dynamics simulations and quantum mechanical current calculations to examine the feasibility of this approach. We find that if some control is exerted over the DNA dynamics, the {\em distributions} of current values for each nucleotide will be sufficiently different to allow for rapid sequencing. We show that a transverse field of the same magnitude as that driving the current provides sufficient control.

\begin{figure}
\includegraphics*[width=7.5cm]{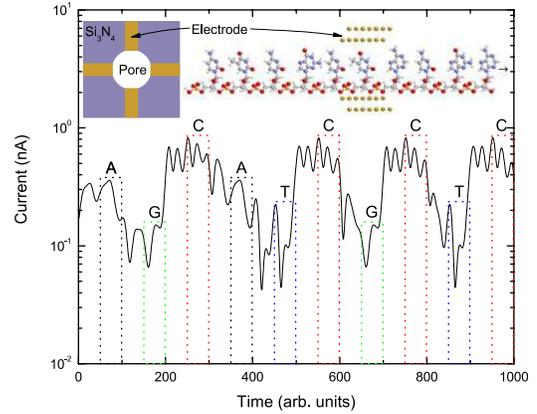}
\caption{Transverse current versus time (in arbitrary units) of a
highly idealized single strand of DNA translocating through a
nanopore with a constant motion. The sequence of the single strand
is AGCATCGCTC. The left inset shows a top-view schematic of the pore
cross section with four electrodes (represented by gold rectangles).
The right inset shows an atomistic side view of the idealized single
strand of DNA and one set of gold electrodes across which electrical
current is calculated. The boxes show half the time each nucleotide
spends in the junction. Within each box, a unique signal from each
of the bases can be seen. } \label{schematic}
\end{figure}

We first discuss an idealized case of DNA dynamics which sets the foundations for the approach we describe. Second, we look at the
distributions of transverse currents through the nucleotides in a realistic setting using a combination of quantum-mechanical
calculations of current and molecular dynamics simulations of DNA translocation through the pore. We use a Green's function method to calculate the current across the electrodes embedded in the nanopore, as described in Ref.~\cite{ZwolakNanolett05}. A tight-binding model is used to represent the molecule and electrode gold atoms. For each carbon, nitrogen, oxygen, and phosphorus atom $s-, p_x-, p_y-$, and $p_z-$ orbitals are used, while $s$-orbitals are used for hydrogen and gold. The retarded Green's function, $\mathcal{G}_{DNA}$, of the system can then be written as
\begin{equation}
\mathcal{G}_{DNA}(E)=[E\mathcal{S}_{DNA}-\mathcal{H}_{DNA}-\Sigma_t-\Sigma_b]^{-1},
\end{equation}
where $\mathcal{S}_{DNA}$ and $\mathcal{H}_{DNA}$ are the overlap and the Hamiltonian matrices~\cite{Yaehmop1}. $\Sigma_{t(b)}$ are the self energy terms describing the coupling between the electrodes and the DNA. The total current can then be expressed as
\begin{equation}
I = \frac{2 e}{h} \int_{-\infty}^{\infty} dE T(E) [f_t(E) -f_b(E)],
\end{equation}
where $T(E)$ is the transmission coefficient and is given by
\begin{equation}
T(E)=\mathrm{Tr}[\Gamma_t\mathcal{G}_{DNA}\Gamma_b\mathcal{G}_{DNA}^\dagger].
\end{equation}
$f_{t(b)}$ is the Fermi-Dirac function of top (bottom) electrode, and $\Gamma_{t(b)}=i(\Sigma_{t(b)}-\Sigma_{t(b)}^\dagger)$. The electrodes are comprised of 3x3 gold atoms arranged as a (111) surface two layers deep, and are biased at 1 V. The electrode spacing is 12.5 \AA. Room temperature has been used for all calculations throughout the paper.

The first question is whether it is at all possible, in the best case scenario, to see differences in the transverse current between the different nucleotides in the absence of structural fluctuations, ions, and water. We address this by studying a highly idealized case of DNA translocation dynamics. The transverse current of a random sequence of single-stranded DNA (ss-DNA) moving through the junction with a constant motion is shown in Figure 1. This figure shows that the different nucleotides do indeed have unique electronic signals in this
ideal case. Similar results have been obtained for static configurations of nucleotides in a previous theoretical work by two of the present authors~\cite{ZwolakNanolett05}, where, in addition, it was shown that neighboring bases do not affect the electronic signature of a given base so long as the electrode widths are of nanometer scale, i.e., of the order of the base spacing. These results provide a good indication that DNA can be sequenced if its dynamics through the pore can be controlled. As we show below, such control is provided by a transverse field of the same magnitude as that driving the current.

Obviously, in a real device there will always be fluctuations of the
current. These fluctuations are mainly due to two sources: 1)
structural fluctuations of the DNA, ions and water, and 2) noise
associated with the electrical current itself, like thermal, shot
and 1/f noise~\cite{LagerqvistNanotech04}. Apart from 1/f noise,
which can be overcome by operating slightly away from the
zero-frequency limit, we estimate that, for the case at hand, shot
noise and thermal noise are negligibly small, giving rise to less
than 0.1\% of error in the current~\cite{LagerqvistNanotech04,prec}.
The most significant source of noise is thus due the structural
motion of the DNA and its environment~\cite{longitudinal}.

We have explored this structural noise by coupling molecular
dynamics simulations with electronic transport calculations
(described above) to obtain the real-time transverse current of the
ss-DNA translocating through a $\textrm{Si}_3\textrm{N}_4$
nanopore~\cite{moldyn}. The $\textrm{Si}_3\textrm{N}_4$ making up
the membrane is assumed to be in the  $\beta$-phase~\cite{si3n4}
with funnel-like shape (see Figure 2), while the electrodes are
described above. A larger distance only reduces the current, while a
shorter distance does not allow easy translocation of the DNA. As we
describe below, the actual geometry of the electrodes and pore does
not change the protocol we suggest for sequencing. The positions of
the atoms of the nanopore and electrodes are assumed to be frozen
throughout the simulation. The electric field generated by the
electrodes is not included when the ss-DNA translocates through the
pore, since the driving field is much larger in magnitude. Its
effect will be analyzed later. A large driving field of 10 kcal/(mol
$\textrm{ \AA}$ e) is used to achieve feasible simulation times. In
experiments such a large field would not be necessary.

For convenience we choose to study the current that flows across two
pairs of mutually-opposite electrodes (see inset of Figure 1). The
four electrodes are not necessary for the conclusions we draw (in an
experiment two are enough~\cite{TaoJphyschem93}). However, analyzing
the current in two perpendicular directions gives us additional
information on the orientation of a nucleotide inside the pore. For
instance, if the ratio between the two currents is large, we know
the nucleotide is aligned in the direction of the electrodes with
the larger current.  If the two currents are about equal in both
directions, it is likely that the base is aligned at a 45 degree
angle, and so forth. This is illustrated by the snapshots in Figure
2 where we see the expected behavior of the current for an ss-DNA
with fifteen consecutive Cytosine bases translocating between the
two pairs of electrodes~\cite{inelastic}.

\begin{figure}
\includegraphics*[width=7.5cm]{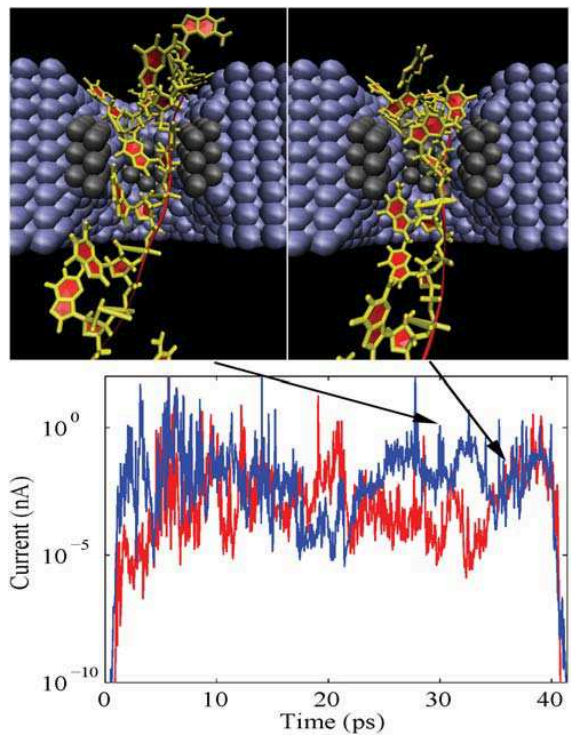}
\caption{Currents as a function of time for a
$\mathrm{poly(dC)}_\mathrm{15}$ translocating through a nanopore.
Blue (red) curve indicates the current, for a bias of 1 V, between
the right and left (front and back) electrodes represented in gray
in the snapshots (the fourth electrode is located behind the field
of view and is hence not visible in the snapshots). During
approximately the first half of the translocation, the two currents
follow each other, indicating no bases are aligned with either
electrode pair. Left snapshot indicates the case in which a
nucleotide is aligned with a pair of electrodes; the right snapshot
when the nucleotide is not aligned between either pair of
electrodes. In the snapshots, solution atoms are not shown and red
colors are a guide for the eye only.} \label{schematic}
\end{figure}

We have found similar curves for all other bases as well, making it difficult to sequence DNA on the basis of just a simple read-out of the current, like what these curves show. In other words, due to structural fluctuations and the irregular dynamics of the ss-DNA, a single measurement of the current for each base is not enough to distinguish the different bases with high precision
(see also Supporting Material). We thus conclude that a {\it distribution} of electrical current values for each base needs to be obtained. This can be done by slowing the DNA translocation in the pore~\cite{FologeaNanolett052} so that each base spends a larger amount of time aligned with the electrodes. Most importantly, we find that when the field that drives the DNA through the pore is smaller than the transverse field that generates the current, one base at a time can align with a pair of electrodes quite easily. This is due to the fact that the DNA backbone is charged in solution, so that its position can be controlled by the transverse field (see also Supporting Material).

Figure 3 shows the main results of this Letter. It shows the calculated distribution of transverse currents for each base in a realistic setting when the driving field is much smaller than the transverse field. We obtain these distributions by turning off the driving field and sampling the current while one base fluctuates between the electrodes~\cite{distributions}. The distributions for each base are indeed different. Note that these distributions may vary according to the microscopic geometry of the pore and electrodes, but our suggested protocol to sequence via transverse transport remains the same. {\em First}, one needs to ``calibrate'' a given nanopore device by obtaining the distributions of current with, say, short homogeneous polynucleotides, one for each base. {\em Second}, once these ``target'' distributions are obtained, a given sequence can be extracted with the {\it same} device by comparing the various currents with these ``target'' distributions, and thus assigning a base to each measurement within a certain statistical accuracy. Both the target and sequencing distributions need to be obtained under the conditions we have discussed above, i.e. the driving field smaller than the transverse field, which allow the transverse field to control the nucleotides alignment with respect to the electrodes.

\begin{figure}
\begin{center}
\includegraphics*[width=7.5cm]{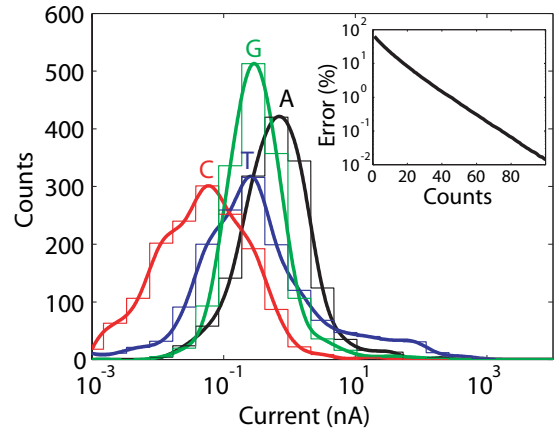}
\caption{Probability distributions of currents at a bias of 1 V for
$\mathrm{poly(dX)}_\mathrm{15}$, where X is
Adenine/Thymine/Cytosine/Guanine for the black/blue/red/green curve,
respectively. The thin lines show the actual current intervals used for the count, while the thick lines are an
interpolation. The inset shows the exponentially decaying ratio of falsely identified bases versus number of independent counts
(measurements) of the current averaged over the four bases. }
\label{schematic}
\end{center}
\end{figure}

Finally, given these distributions and the accuracy with which we want to sequence DNA, we can answer the question of how many independent electrical current measurements one needs to do in order to sequence DNA within that accuracy. The number of current measurements will dictate how fast we can sequence. We can easily estimate this speed from the distributions of Figure 3 by calculating the statistical likelihood for all configurations of a given base in the junction region and multiplying it by the probability that we can tell this base from all other bases for the value of the current at that specific configuration. The average probability that we can correctly sequence a base after N measurements is then given by
\begin{equation}
\begin{split}
<P>&=\sum_{X=A,T,C,G}\frac{1}{4} \sum_{\{I_n\}}\\
&\frac{\prod_{n=1}^{N} P_X^n}{\prod_{n=1}^{N} P_A^n+\prod_{n=1}^{N} P_T^n+\prod_{n=1}^{N} P_C^n+\prod_{n=1}^{N} P_G^n}
\end{split}
\end{equation}
where A, T, C and G are the distributions, as shown in Figure 3, for the four bases. $P_X^n$ is the probability that a base is X considering only the current for measurement $n$. It can be found by comparing the ratios of the four distributions. Finally, the sum over $\{I_n\}$ is a sum over all possible sets of measurements of size $N$. The inset of Figure 3 shows $1-<P>$, the exponentially decaying ratio of falsely identified bases versus number of independent counts (measurements) of the current averaged over the four bases, where the ensemble average is performed using Monte-Carlo methods. From this inset we see that if, for instance, we want to sequence DNA with an error of 0.1\%, we need about 80 electrical current measurements to distinguish the four bases. If we are able to collect, say, $10^7$ measurements of the current per second (a typical rate of electrical current measurements) we can sequence the whole genome in less than seven hours without parallelization. Note that it is mainly the rate at which electrical current measurements can be done that sets an upper limit for the sequencing speed, not the DNA translocation speed. Clearly, these estimates may vary with different device structures but are representative of the speeds attainable with this sequencing method.

We thus conclude that the approach we have described in this Letter
shows tremendous potential as an alternative sequencing method.  If
successfully implemented, DNA sequencing could be performed orders
of magnitude faster than currently available methods and still much
faster than other pre-production approaches recently
suggested~\cite{CollinsNat03,ShendureNatgen04,MarguliesNat05}.

{\bf Acknowledgments} We gratefully acknowledge discussions with M. Ramsey and T. Schindler. This research is supported by the NIH-National Human Genome Research Institute (JL and MD) and by the National Science Foundation through its Graduate Fellowship program (MZ).

{\bf Supporting Information Available} We include subsidiary material which contains two movies, one showing the translocation dynamics and the other the control exerted by the transverse field.  This material is available
free of charge via the Internet at http://pubs.acs.org.


\begin{thebibliography}{99}

\bibitem[*]{MD}E-mail address: diventra@physics.ucsd.edu.

\bibitem{LiNat01} Li, J.; Stein, D.; McMullan, C.; Branton, D.; Aziz, M. J.; Golovchenko, J.A.
{\it Nature} {\bf 2001}, {\it 412}, 166.

\bibitem{StormNatmat03} Storm, A. J.; Chen, J. H.; Ling, X. S.; Zandbergen, H. W.; Dekker, C.
{\it Nature Mat.} {\bf 2003}, {\it 2}, 537.

\bibitem{HarrellAnalchem03} Harrell, C. C.; Lee, S. B.; Martin, C. R.
{\it Anal. Chem.} {\bf 2003}, {\it 75}, 6861.

\bibitem{LiAnalchem04} Li, N.; Yu, S.; Harrell, C. C.; Martin, C. R.
{\it Anal. Chem.} {\bf 2004}, {\it 76}, 2025.

\bibitem{LemayAnalchem05} Lemay, S. G; van den Broek, D. M.; Storm, A. J.; Krapf, D.; Smeets, R. M. M.; Heering, H. A.; Dekker, C.
{\it Anal. Chem.} {\bf 2005}, {\it 77}, 1911.

\bibitem{KasianowiczProcusa96} Kasianowicz, J. J.; Brandin, E.; Branton, D.; Deamer, D. W.
{\it Proc. Natl. Acad. Sci. USA } {\bf 1996}, {\it 93}, 13770.

\bibitem{AkesonBiophys99} Akeson, M.; Branton, D.; Kasianowicz, J. J.; Brandin, E.; Deamer, D.W.
{\it Biophys. J.} {\bf 1999}, {\it 77}, 3227.

\bibitem{DeamerTrends00} Deamer, D. W.; Akeson, M.
{\it Trends in Biotechnology} {\bf 2000}, {\it 18}, 147.

\bibitem{MellerProcusa00} Meller, A.; Nivon, L.; Brandin, E.; Golovchenko, J.; Branton, D.
{\it Proc. Natl. Acad. Sci. USA} {\bf 2000}, {\it 97}, 1079.

\bibitem{VercoutereNatbio01} Vercoutere, W.; Winters-Hilt, S.; Olsen, H.; Deamer, D.; Haussler, D.; Akeson, M.
{\it Nature Biotechnology} {\bf 2001}, {\it 19}, 248.

\bibitem{MellerPRL01} Meller, A.; Nivon, L.; Branton, D.
{\it Phys. Rev. Lett.} {\bf 2001}, {\it 86}, 3435.

\bibitem{DeamerAccchem02} Deamer, D.; Branton, D.
{\it Acc. Chem. Res.} {\bf 2002}, {\it 35}, 817.

\bibitem{MellerElectro02} Meller, A.; Branton, D.
{\it Electrophoresis} {\bf 2002}, {\it 23}, 2583.

\bibitem{LiNatmat03} Li, J.; Gershow, M.; Stein, D.; Brandin, E.; Golovchenko, J. A.
{\it Nature Mat.} {\bf 2003}, {\it 2}, 611.

\bibitem{NakaneJphyscm03} Nakane, J. J.; Akeson, M.; Marziali, A.
{\it J. Phys. Cond. Matt.} {\bf 2003}, {\it 15}, R1365.

\bibitem{AksimentievBiophys04} Aksimentiev, A.; Heng, J. B.; Timp, G.; Schulten, K.
{\it Biophys. J.} {\bf 2004}, {\it 87}, 2086.

\bibitem{ChenNanolett04} Chen, P.; Gu, J.; Brandin, E.; Kim, Y.-R.; Wang, Q.; Branton, D.
{\it Nano Lett.} {\bf 2004}, {\it 4}, 2293.

\bibitem{FologeaNanolett051} Fologea, D. ; Gershow, M.; Ledden, B.; McNabb, D. S.; Golovchenko, J. A.; Li, J.
{\it Nano Lett.} {\bf 2005}, {\it 5}, 1905.

\bibitem{HengBiophys06} Heng, J. B.; Aksimentiev, A.; Ho, C.; Marks, P.; Grinkova, Y. V.; Sligar, S.; Schulten, K.; Timp, G.
{\it Biophys. J.} {\bf 2006}, {\it 90}, 1098.


\bibitem{RefAMIT} Braslavsky, I.; Hebert, B.; Kartalov, E.; Quake, S. R.
{\it Proc. Natl. Acad. Sci. USA} {\bf 2003}, {\it 100}, 3960.

\bibitem{RefSchulten06} Gracheva, M. E.; Xiong, A.; Aksimentiev, A.; Schulten, K.; Timp, G.; Leburton,
J.-P.
{\it Nanotechnology} {\bf 2006}, {\it 17}, 622.

\bibitem{ChanMutat05} Chan, E. Y.
{\it Mutat. Res.} {\bf 2005}, {\it 573}, 13.

\bibitem{ZwolakNanolett05} Zwolak, M.; Di Ventra, M.
{\it Nano Lett.} {\bf 2005}, {\it 5}, 421.

\bibitem{Yaehmop1} We have used the program YAeHMOP (http://yaehmop.sourceforge.net/) to generate the tight-binding parameters used in this work. These parameters reproduce the states near the highest occupied and the lowest unoccupied molecular orbitals quite well compared to density-functional (DFT) calculations on the passivated nucleotide. In solution though, the nucleotides are charged. We have performed DFT calculations on charged nucleotides and found that the complex chemical environment with water molecules, counter-ions, and the gold electrodes effectively passivates the charge on the nucleotides. This finding allows us to use the same tight-binding parameters as in the passivated case. On the other hand, water only lowers the electrical current by a few percent on average, and its effect on the distributions is negligible. We will discuss these issues in more detail in a future publication.

\bibitem{LagerqvistNanotech04} Lagerqvist, J.; Chen, Y.-C.; Di Ventra, M.
{\it Nanotechnology} {\bf 2004}, {\it 15}, S459; Chen, Y.-C.; Di
Ventra, M. {\it Phys. Rev. Lett.} {\bf 2005}, {\it 95}, 166802.

\bibitem{prec} For instance, for a typical current of 1nA and the bias of 1V we consider in this work, if we assume a frequency
bandwidth $\Delta f$ of 10 kHz, the thermal noise can be estimated to be $i_n=\sqrt{\frac{4k_B T \Delta f}{R}}=0.4 \textrm{pA}$ at room temperature. The effect of shot noise is comparable or less.

\bibitem{longitudinal} These effects are also very important for
charge transport in DNA along the longitudinal direction, see, e.g.,
Di Ventra, M.; Zwolak, M. DNA Electronics. \textit{Encyclopedia of
Nanoscience and Nanotechnology}, Singh-Nalwa H., Eds.; American
Scientific Publishers, 2004, Vol. 2, p. 475; Endres, R. G.; Cox, D.
L.; Singh, R. R. P. {\it Rev. Mod. Phys.} {\bf 2004}, {\it 76}, 195;
Porath D.; Cuniberti, G.; Di Felice, R. {\it Topics in Current
Chemistry} {\bf 2004}, {\it 237}, 183.

\bibitem{moldyn}The dynamics of ss-DNA in the pore was calculated using the classical molecular dynamics package NAMD2~\cite{namd2}. For the DNA, TIP3 water, and ions (1 M KCl) the CHARMM27~\cite{charmm271,charmm272} force field was used. UFF parameters~\cite{UFF} were used for the interaction between the device itself and other elements.

\bibitem{namd2} Kale L.; Skeel, R.; Bhandarkar, M.; Brunner, R.; Gursoy, A.; Krawetz, N.; Phillips, J.; Shinozaki, A.; Varadarajan, K.; Schulten, K.
{\it J. Comp. Phys.} {\bf 1999}, {\it 151}, 283.

\bibitem{charmm271} Foloppe, N.; MacKerell Jr., A. D.
{\it J. Comp. Chem.} {\bf 2000}, {\it 21}, 86.

\bibitem{charmm272} MacKerell Jr., A. D.; Banavali, N. K.
{\it J. Comp. Chem.} {\bf 2000}, {\it 21}, 105.

\bibitem{UFF} Rappe, A. K.; Casewit, C. J.; Colwell, K. S. Goddard III, W. A.; Skiff, W. M.
{\it J. Am. Chem. Soc.} {\bf 1992}, {\it 114}, 10024.

\bibitem{si3n4} Grun, R.
{\it Acta Crystallogr. B} {\bf 1979}, {\it 35}, 800.

\bibitem{TaoJphyschem93} Tao, N. J.; DeRose, J. A.; Lindsay, S. M.
{\it J. Phys. Chem.} {\bf 1993}, {\it 97}, 910.

\bibitem{inelastic} We stress again that for the conclusions of our paper we are only interested in the current of mutually-opposite electrodes. In any case, the current in between two neighboring electrodes would require inelastic scattering events, and hence is expected to be small.

\bibitem{FologeaNanolett052} Fologea, D.; Uplinger, J.; Thomas, B.; McNabb, D. S.; Li, J.
{\it Nano Lett.} {\bf 2005}, {\it 5}, 1734.

\bibitem{distributions} While collecting these distributions, the transverse field of strength 1.84 kcal/(mol $\textrm{\AA}$  e) which generates the current is on during the molecular dynamics simulations. During a time span of 75 ps, we record 1500 values of the current. The switch-off of the driving field is dictated by the molecular dynamics timescales: we cannot slow down the translocation to an experimentally accessible regime.

\bibitem{CollinsNat03} Collins, F. S.; Green, E. D.; Guttmacher, A. E.; Guyer, M. S.
{\it Nature} {\bf 2003}, {\it 422}, 835.

\bibitem{ShendureNatgen04} Shendure, J.; Mitra, R. D.; Varma, C.; Church, G. M.
{\it Nature Reviews Genetics} {\bf 2004}, {\it 5}, 335.

\bibitem{MarguliesNat05} Margulies, M. et al.
{\it Nature} {\bf 2005}, {\it 437}, 376.

\end{thebibliography}
\end{document}